\documentclass[12pt]{article}

\usepackage[T2A]{fontenc}
\usepackage[utf8]{inputenc}
\usepackage[english]{babel}
\usepackage{amsmath}
\usepackage{amsfonts}
\usepackage{mathrsfs}
\usepackage{amssymb}
\usepackage{cite}

\def\theequation{\arabic{section}.\arabic{equation}}
\numberwithin{equation}{section}

\newcommand{\be}{\begin{equation}}
\newcommand{\ee}{\end{equation}}
\newcommand{\bea}{\begin{eqnarray}}
\newcommand{\eea}{\end{eqnarray}}

\newcommand{\p}[1]{(\ref{#1})}

\begin{document}

\begin{titlepage}

\vspace*{0.7cm}

\begin{center}

{\LARGE\bf Twistor formulation of}

\vspace{0.5cm}

{\LARGE\bf massless $6D$ infinite spin fields}

\vspace{1.5cm}

{\large\bf I.L.\,Buchbinder$^{1,2,3}$\!\!,\ \ \  S.A.\,Fedoruk$^3$\!\!,\ \ \ A.P.\,Isaev$^{3,4}$}

\vspace{1.0cm}

\ $^1${\it Center of Theoretical Physics,
Tomsk State Pedagogical University, \\
634041 Tomsk, Russia}, \\
{\tt joseph@tspu.edu.ru}

\vskip 0.3cm

\ $^2${\it National Research Tomsk State  University,}\\{634050 Tomsk, Russia}

\vskip 0.3cm

\ $^3${\it Bogoliubov Laboratory of Theoretical Physics,
Joint Institute for Nuclear Research, \\
141980 Dubna, Moscow Region, Russia}, \\
{\tt fedoruk@theor.jinr.ru, isaevap@theor.jinr.ru}

\vskip 0.3cm

\ $^4${\it St.Petersburg Department of Steklov Mathematical
Institute of RAS, \\ 191023 St. Petersburg, Russia}

\end{center}

\vspace{1.5cm}

\begin{abstract}
We construct massless infinite spin irreducible
representations of the six-dimensional Poincar\'{e} group in the space of
fields depending on twistor variables. It is shown that the massless
infinite spin representation is realized on the two-twistor fields.
We present a full set of equations of motion for two-twistor fields
represented by the totally symmetric $\mathrm{SU}(2)$ rank $2s$
two-twistor spin-tensor and show that they carry massless
infinite spin representations.
A field twistor transform is constructed and infinite spin fields are found in the space-time formulation
with an additional spinor coordinate.
\end{abstract}

\vspace{1.5cm}

%\bigskip

\phantom{Keywords: }

\vspace{1cm}

\end{titlepage}

\setcounter{footnote}{0}
\setcounter{equation}{0}

\newpage

\setcounter{equation}0
\section{Introduction}

Recently, there has been a certain interest in the study of
representations of the Poincar\'{e} group in the higher dimensional
Minkowski space (see e.g. \cite{Weinberg,Kuzenko,BFIP} and
references therein). There are at least two main motivations for
research in space-time symmetry of multidimensional space-time. One
is related to the low-energy limits of the superstring theory which
are supersymmetric gauge theories and supergravity in ten
dimensions. After reduction we can obtain the supersymmetric
field models in all dimensions less than ten where many of the specific
details, in particular the properties of fermionic fields,  are
due to space-time symmetries. The other motivation is rather formal
and is stipulated by the possibilities to generalize the methods for
describing irreducible representations of the Poincar\'{e} group
in four dimensions \cite{Wigner39,Wigner47,BargWigner} up to higher
dimensions (see e.g. \cite{BekMou,BekBoul,BekBoul1}). This can be
useful for studying various aspects of the higher spin field theory
in higher dimensions.

In the paper \cite{BFIP}, we presented a general description of
massless irreducible representations of the Poincar\'{e} group in
the six-dimensional Minkowski space. The corresponding Lie algebra
possesses three Casimir operators whose eigenvalues were
found. It was shown that finite spin representations are defined
by two integer or half-integer numbers while infinite spin
representations are defined by the real parameter $\mu^2$ and one
integer or half-integer number. One of the most important
applications of the Poincar\'{e} group is due to field theory where
we need representations in the space of fields. Therefore, our
next problem is to construct a realization of the general
description \cite{BFIP} on the fields in six-dimensional space-time.
The paper under consideration is devoted to aspects of this problem
related to infinite spin representations. To be more precise, we
will consider the realization of massless representations in the
space of twistor fields.

We will focus on describing massless infinite spin
irreducible twistor field representations in the six-dimensional
Minkowski space. In recent years, there has been a surge of activity
in the study of various aspects of fields with infinite spin,
mainly in the context of higher spin field theory (see e.g. the review
\cite{BekSk} and refs. therein; see also
\cite{Bekaert:2017xin,Najafizadeh:2017tin,HabZin,AlkGr,Metsaev18,BFIR,BFI,BuchKrTak,BuchIFKr,ACG18,
Metsaev18a,Metsaev19,BKSZ,MN20}).
In this paper, we are going to discuss
issues related to the realization of massless infinite spin
representations on twistor fields in the six-dimensional
Minkowski space.

Approaches to the twistor description of massless finite spin
(super)particles and fields in six dimensions were discussed in
Refs. \cite{BeBeCeLi,BeCe,DGalS,GalHS,GalHT,MezRTown14,AMezTown}. In
this paper, we will study the twistor field formulation of infinite
spin representations in six dimensions as a generalization of the
previously proposed twistor formulation of an infinite spin  particle
in four dimensions \cite{BFIR,BFI}. As far as we know, these aspects
were not considered in detail earlier.

The paper is organized as follows. Section 2 is devoted to a brief
survey of the irreducible massless representations of the
Poincar\'{e} group in the $D{=}\,6$ Minkowski space. In Section 3, we consider
one-twistor space and construct a realization of the Poincar\'{e}
group generators in the space of one-twistor fields. We show that
this space necessarily describes finite spin representations and
its generalization is required to solve our problems related to
infinite spin representations.
Section 4 is devoted to the implementation
of massless representations with infinite spin. We introduce two-twistor space and
two-twistor fields, describe the realization of the generators in
the space of these fields and show that these fields form
infinite spin representations. In Section 5, infinite spin
representations are constructed in terms of totally symmetric $2s$
rank spin-tensor twistor fields. In Section 6, we construct a twistor transform
that links the found twistor fields with space-time infinite spin fields
in the formulation with an additional spinor variable.
In Section 7, we summarize the
results and discuss open issues. Some technical points related
to the notation and calculation details are referred to Appendices.

\setcounter{equation}0
\section{Massless representations in $D{=}\,6$ Minkowski space-time}

In this section, to fix the notation, we recall the general
results on the irreducible massless representations of the
Poincar\'{e} group in six dimensions \cite{BFIP}.  This description
will be used in the next sections for the construction of this type of
representations in the space of twistor fields.

Relativistic symmetry transformations in the six-dimensional Minkowski
space are generated by elements of the Lie algebra
$\mathfrak{iso}(1,5)$ of the corresponding Poincar\'{e} group.
The basis elements $P_{m}$ and $M_{mn}=-M_{nm}$ of
$\mathfrak{iso}(1,5)$ have the commutators
\begin{equation}
\label{P-M}
[P_{n},P_{k}]= 0 \; , \;\;\;\;\;
[M_{mn},P_{k}]=i\left(\eta_{mk}P_{n}-\eta_{nk}P_{m}\right)\,,
\end{equation}
\begin{equation}
\label{M-M}
[M_{mn},M_{kl}]=i\left(\eta_{mk}M_{nl}+\eta_{nl}M_{mk}-\eta_{ml}M_{nk}
-\eta_{nk}M_{ml}\right)\,.
\end{equation}
We use the space-time metric $\eta_{mn}={\rm
diag}(+1,-1,-1,-1,-1,-1)$ and vector indices $m,n,...$ run through six
values $0,1,\ldots,5$.

The quadratic Casimir operator of the Lie algebra $\mathfrak{iso}(1,5)$ has the form
\begin{equation}
\label{P2-def}
C_2 \ = \ P^2 \ = \ P^{m}P_{m} \,.
\end{equation}
When acting on the states of an irreducible representation, the operator \p{P2-def} takes a fixed numerical value
equal to the square of a massive parameter which in some
 cases is interpreted as
a relativistic particle mass.
Since we consider massless representations,
the Casimir operator \p{P2-def} is equal to zero:
\begin{equation}
\label{P2-0}
C_2 \ = \ 0\,.
\end{equation}
In this case, when we have \p{P2-0},
the other two Casimir operators of
the Lie algebra $\mathfrak{iso}(1,5)$  are given
by the following expressions \cite{BFIP}:
\begin{eqnarray}
\label{C4-Cas}
C_4&=&  \Pi^{m} \Pi_{m} \,, \\ [7pt]
\label{C6-Cas}
C_6&=&
-\, \Pi^k M_{km}\, \Pi_l M^{lm}
\ + \ \frac12\,\Big(M^{mn}M_{mn}-8\Big)\, C_4 \,,
\end{eqnarray}
where
\begin{equation}
\label{Pi-def}
\Pi_m := P^{k}  M_{km}  \; .
\end{equation}

Massless finite spin representations (so called
helicity representations) are defined in the space of fields
where the Casimir operators \p{C4-Cas} and
\p{C6-Cas} have zero eigenvalues. They are characterized by two spin (helicity) operators,
the explicit expressions of which are given in \cite{BFIP}.

The massless infinite (continuous) spin representations are realized in the space of states on which
the operators \p{C4-Cas}, \p{C6-Cas} acquire numerical values:
\begin{eqnarray}
\label{C4-ir}
C_4&=&  -\mu^2 \,, \\ [7pt]
\label{C6-ir}
C_6&=& -\mu^2 s(s+1)\,,
\end{eqnarray}
where $\mu \neq 0$ is the dimensionful real parameter
 (we assume that $\mu \in
\mathbb{R}_{> 0}$) and $s$ is a non-negative integer or
half-integer number $s \in \mathbb{Z}_{\geq 0}/2$. Of course, in
the limit $\mu{=}\,0$ the finite spin (helicity) representations
are reproduced.

In the following sections, we will find a twistor representation
of the algebra \p{P-M}, \p{M-M} and study fields
in twistor space where massless infinite spin representations
are explicitly realized.

\setcounter{equation}0
\section{One-twistor case: finite spin representations}

Let us first consider the one-twistor case.

Following the standard prescription of the twistor formalism (see,
e.g., \cite{BeBeCeLi,BeCe,MezRTown14,AMezTown}), we consider a
twistor as an object consisting of two chiral spinors,
which are canonically conjugate to each other
and thus form the phase space of classical mechanics.

In $D\,{=}\,(1{+}5)$
Minkowski space-time there are no standard
Majorana-Weyl spinors but there are
$\mathrm{SU}(2)$ ``real'' spinors
\footnote{We call them the $\mathrm{SU}(2)$ Majorana-Weyl spinors.}
\cite{KuTow} (see also \cite{IR}, Sect.\,6.4.2)
which in general can have different
chiralities.
So we take the $\mathrm{SU}(2)$ Majorana-Weyl spinor with fixed chirality
\begin{equation}
\label{pi} \pi_\alpha^I\,, \qquad (\pi_\alpha^I)^* =
\epsilon_{IJ}B_{\dot\alpha}{}^\beta \pi_\beta^J,
\end{equation}
where $B_{\dot\alpha}{}^\beta$ is the conjugation matrix.
Here $\alpha=1,2,3,4$ is the $\mathrm{Spin}(1,5)\simeq
\mathrm{SU}^*(4)$ index and $I,J=1,2$ are the $\mathrm{SU}(2)$
indices,\footnote{Throughout this paper, we raise and lower the
$\mathrm{SU}(2)$ indices as follows: $\psi^I=\epsilon^{IJ}\psi_J$,
$\phi_I=\epsilon_{IJ}\psi^J$ where the antisymmetric tensors
$\epsilon_{IJ}$, $\epsilon^{IJ}$ have components
$\epsilon_{12}=\epsilon^{21}=1$. The details of our spinor
notation are given in Appendix\,A.} and consider this spinor as half of the $D{=}\,(1+5)$
twistor. The second half of the twistor is the $\mathrm{SU}(2)$
Majorana-Weyl spinor
\begin{equation}
\label{omega}
\omega^{\alpha I}\,, \qquad (\omega^{\alpha I})^* = \epsilon_{IJ}\omega^{\beta J}(B^{-1})_\beta{}^{\dot\alpha}
\end{equation}
having the opposite to the spinor \p{pi}
chirality. The basic twistor Poisson brackets have the form
\begin{equation}
\label{PB-tw}
\left\{ \pi_\alpha^I\,, \omega^\beta_J\right\}_{PB} = \delta_\alpha^\beta \delta^I_J \,.
\end{equation}
Thus, the $D{=}\,(1+5)$ twistor consists of
two $\mathrm{SU}(2)$ Majorana-Weyl spinors  \p{pi}, \p{omega}
\begin{equation}
\label{Z-tw}
Z_a^I=\left(
\begin{array}{c}
\!\!\!\!\pi_\alpha^I \\ [4pt]
\omega{}^\beta{}^I \\
\end{array}
\!\!\!\right),\qquad
\omega{}^\beta{}^I=\epsilon^{IJ}\omega{}^\beta_J\,,
\end{equation}
where the index $a$ runs through eight values $a=1,\ldots,8$.
The quantities
\begin{equation}
\label{6-conf}
X_{[ab]}:=Z_a^I Z_b^J \epsilon_{IJ}
\end{equation}
form the $\mathfrak{so}(2,6)\simeq \mathfrak{so}^*(8)$ algebra with respect to the Poisson brackets \p{PB-tw} and
generate local $\mathrm{Spin}(2,6)$ transformations as linear transformations of the twistor \p{Z-tw} (see, for example, \cite{HuLip}
and Appendix\,B).
So, as in the standard twistor prescription \cite{MezRTown14,AMezTown},
the twistor\p{Z-tw} is a couple of
the $\mathrm{SU}(2)$ Majorana-Weyl spinors, which
form generators \p{6-conf}
 of the $D{=}\,(1+5)$ conformal group $\mathrm{Spin}(2,6)$.
We emphasize that $Z^I$ for $I=1$ and $I=2$
are related by the reality conditions
\p{pi}, \p{omega}. Thus, despite the presence
of the $\mathrm{SU}(2)$ index $I=1,2$,
we can talk about only {\it one complex twistor}.

Below we will consider the field twistor theory where the Poisson
brackets \p{PB-tw} are quantized and become commutators
\begin{equation}
\label{com-tw}
\left[ \pi_\alpha^I\,, \omega^\beta_J\right] = i\delta_\alpha^\beta \delta^I_J
\end{equation}
of the operators acting on the twistor fields.
Then we choose the $\pi$-representation for
the twistor fields in which the coordinates $\pi_\alpha^I$ are ``diagonal'',
one-twistor fields $\Psi^{(1)}(\pi)$ are the functions of these coordinates,
and the operators $\omega^\alpha_I$ are realized by differential operators
\begin{equation}
\label{d-pi}
\omega^\alpha_I=-i\,\frac{\partial}{\partial\pi_\alpha^I}\,.
\end{equation}

One of the basic properties of the twistor formulation is the
representation of the relativistic momentum vector of a massless
particle in the form of a bilinear combination of twistor coordinates
where the light-like condition \p{P2-0} of the relativistic momentum is
satisfied automatically. Therefore, in the considered one-twistor
case we take the generators of the Poincar\'{e} translations in the
following form:
\begin{equation}
\label{P-1}
P_m = \pi_\alpha^I (\tilde\sigma_{m})^{\alpha\beta}\pi_\beta{}_I\,.
\end{equation}
In fact,  the momentum vector \p{P-1} is light-like due to
the property \p{sigma-eq3}:
$$
P^mP_m =2\epsilon^{\alpha\beta\gamma\delta}\pi_\alpha^I \pi_\beta{}_I\pi_\gamma^J \pi_\delta{}_J\equiv 0 \,.
$$
Thus, for the representation \p{P-1},
the one-twistor fields  $\Psi^{(1)}(\pi)$ describe massless
states. The generators
$M_{mn}$ of the Lorentz algebra $\mathfrak{so}(1,5)$ act
in the space of fields $\Psi^{(1)}(\pi)$
as operators
\begin{equation}
\label{M-1}
M_{mn} = -i\pi_\alpha^I (\tilde\sigma_{mn})^{\alpha}{}_{\beta}\frac{\partial}{\partial\pi_\beta^I}\,.
\end{equation}
We note that in the one-twistor realization \p{P-1}, \p{M-1} of the
Poincar\'{e} generators the operator $\Pi_m$, which is presented in
the definition \p{C4-Cas} of the four-order Casimir operator and defined in \p{Pi-def}, equals
$$
\Pi_m \ = \ \frac{i}{2}\left(\pi^K\tilde\sigma_{m}\pi_K \right)\pi_\alpha^I\frac{\partial}{\partial\pi_\alpha^I}
\ \equiv \ \frac{i}{2} \, P_m \;
\pi_\alpha^I\frac{\partial}{\partial\pi_\alpha^I} \ .
$$
As a result, in view of \p{P2-0},
the Casimir operator \p{C4-Cas}
vanishes: $C_4=0$, and in this case  $\mu=0$
due to \p{C4-ir}. Then, \p{C6-ir} shows that
in one-twistor case the six-order Casimir
operator is also equal to zero: $C_6=0$.

Thus, the one-twistor field $\Psi^{(1)}(\pi)$ describes only
massless finite spin representations (helicity representations). For
description of massless infinite (continuous) spin
representations it is necessary to use two or more twistors
\cite{MezRTown14} as in
the massive representations in the $D=4$ case (see, e.g., \cite{AFIL,IsPod}).

\setcounter{equation}0
\section{Two-twistor case: twistorial constraints in infinite-spin case}

The massless infinite spin representation of
$\mathfrak{iso}(1,5)$ is formulated
in the space two-twistor fields. To describe these fields, we need
two copies of the twistors that were considered in the previous section.

In addition to the twistor, defined by relations
\p{pi}-\p{Z-tw}, we introduce  exactly in the same way
the second twistor $Y_a^A$:
\begin{equation}
\label{Y-tw}
Y_a^A=\left(
\begin{array}{c}
\!\!\!\!\rho_\alpha^A \\ [4pt]
\eta{}^\beta{}^A \\
\end{array}
\!\!\!\right) \; .
\end{equation}
This twistor consists of the $\mathrm{SU}(2)$ Majorana-Weyl spinor
\begin{equation}
\label{rho}
\rho_{\alpha}^{A} \,, \qquad (\rho_{\alpha}^{A})^* =
\epsilon_{AB}B_{\dot\alpha}{}^\beta \rho_{\beta}^{B} \,,
\end{equation}
which has the same
chirality\footnote{The twistor formulation of the $D{=}\,6$
system, when the upper spinor $\rho_\alpha^A$
of the second twistor $Y^{I}_a$ has the chirality  different from that of the upper spinor $\pi_\alpha^I$ of
the twistor \p{Z-tw}, will be  discussed in the last concluding section.}
as spinor \p{pi}, and another
$\mathrm{SU}(2)$ Majorana-Weyl spinor
\begin{equation}
\label{eta}
\eta^{\alpha A}\,, \qquad (\eta^{\alpha A})^* = \epsilon_{AB}\eta^{\beta B}(B^{-1})_\beta{}^{\dot\alpha}\,,
\end{equation}
which is canonically conjugated to the first half \p{rho}
of the twistor $Y_a^I$.
The nonvanishing  Poisson brackets of the twistor components \p{rho}, \p{eta}
have the form
\begin{equation}
\label{PB-tw2}
\left\{ \rho_\alpha^A\,, \eta^\beta_B\right\}_{PB} = \delta_\alpha^\beta \delta^A_B \, .
\end{equation}
The components \p{rho}, \p{eta} have zero Poisson brackets with variables \p{pi}, \p{omega}.

In expressions \p{Y-tw}-\p{PB-tw2}, $A=1,2$ is the $\mathrm{SU}(2)$ index.
In general, index $I$ in \p{pi}-\p{Z-tw} and index $A$ in \p{Y-tw}-\p{PB-tw2} are different. This means that
the twistors $Z^I$ and $Y^A$ are vectors in the representations
of different $\mathrm{SU}(2)$ groups.\footnote{
In the $D=6$ massive case, the twistor formulation  uses the $\mathrm{USp}(4)$ Majorana-Weyl spinor \cite{MezRTown14,AMezTown}
$$
\pi_\alpha^{\mathcal{I}}=(\pi_\alpha^I,\rho_\alpha^A)\,,
$$
where $\mathcal{I}=1,2,3,4$ is the $\mathrm{USp}(4)$ index. This
$\mathrm{USp}(4)$ Majorana-Weyl spinor is formed by two
$\mathrm{SU}(2)$ Majorana-Weyl spinors $\pi_\alpha^I$ and
$\rho_\alpha^A$. However, in the case of
infinite spin considered here, the
$\mathrm{USp}(4)$ symmetry does not survive and we use the
$\mathrm{SU}(2)$ Majorana-Weyl spinors
separately.}
However, as we will
see, these $\mathrm{SU}(2)$ groups must be identified in the
case of infinite spin representations considered
here. With such identification  the indices
$I$ and $A$ refer to the same symmetry group and
can be contracted to each other.

In the quantum case, the twistor brackets \p{PB-tw2} produce the algebra
\begin{equation}
\label{com-tw2}
\left[ \rho_\alpha^A\,, \eta^\beta_B\right]_{PB} = i\delta_\alpha^\beta \delta^A_B \,.
\end{equation}
Therefore, we can consider together the
$\pi$-representation with realization \p{d-pi} and the $\rho$-representation
in which the quantities $\eta^\alpha_A$ are realized by differential operators
\begin{equation}
\label{d-rho}
\eta^\alpha_A=-i\,\frac{\partial}{\partial\rho_\alpha^A}\,.
\end{equation}
In such representations the physical states are described by the $6D$ two-twistor field
\begin{equation}
\label{Psi}
\Psi=\Psi(\pi_\alpha^I,\rho_\alpha^A)\,,
\end{equation}
which is a function of upper halves
of both twistors \p{Z-tw}, \p{Y-tw}.

Now we construct $6D$-vectors (these vectors will be used
below) by contracting
 spinors $\pi_\alpha^I$, $\rho_\alpha^A$ with the matrices
$\tilde\sigma_{m}$:
\begin{equation}
\label{vectors}
v_m:=\left(\pi^I\tilde\sigma_{m}\pi_I \right)\,, \qquad w_m:=\left(\rho^A\tilde\sigma_{m}\rho_A \right)\,,\qquad
u^{IA}_m:=\left(\pi^I\tilde\sigma_{m}\rho^A \right),
\end{equation}
where contracted spinor indices are omitted:
$\left(\pi^I\tilde\sigma_{m}\rho^A \right)\equiv
\pi_\alpha^I(\tilde\sigma_{m})^{\alpha\beta}\rho_\beta^A$, etc.
Nonvanishing scalar products of the vectors  \p{vectors}
have the form
\begin{equation}
\label{uvw-2}
v^m w_m = r\,,
\qquad u^{m\,IA} u^{JB}_m = -\frac14\,\epsilon^{IJ}
\epsilon^{AB}r\,,
\end{equation}
where
\begin{equation}
\label{r}
r := 2\,\epsilon^{\alpha\beta\gamma\delta} \pi_\alpha^I\pi_\beta{}_I \rho_\gamma^A\rho_\delta{}_A \,.
\end{equation}
Due to identities \p{sigma-eq3}, the other
scalar products of the
vectors \p{vectors} are equal to zero:
\begin{equation}
\label{uvw-0}
v^m v_m=w^m w_m=v^m u^{IA}_m=v^m u^{IA}_m=0\,.
\end{equation}

Since we want to describe the massless states
by the two-twistor fields \p{Psi}, these
 fields must satisfy the massless condition \p{P2-0}.
 This condition is fulfilled automatically if
we take
 the space-time translations $P_m$ in the form \p{P-1},
 which was proposed
in the previous section for the one-twistor case. Thus, only one
of two twistorial spinors $\pi^I_\alpha$ and $\rho^A_\alpha$ resolves
the momentum operators:
\begin{equation}
\label{P-2}
P_m = \pi_\alpha^I (\tilde\sigma_{m})^{\alpha\beta}\pi_\beta{}_I\,.
\end{equation}

In the two-twistor case,
the generators of Lorentz transformations
 act on the spinor components of both twistors and instead of \p{M-1} we have:
\begin{equation}
\label{M-2}
M_{mn}  \ = \
-i\pi_\alpha^I (\tilde\sigma_{mn})^{\alpha}{}_{\beta}\frac{\partial}{\partial
\pi_\beta^I}
-i\rho_\alpha^A (\tilde\sigma_{mn})^{\alpha}{}_{\beta}\frac{\partial}{\partial\rho_\beta^A}\,.
\end{equation}

Using expressions \p{P-2} and \p{M-2}, we obtain that the
vector operator $\Pi_m$, which is defined in
\p{Pi-def}, takes the form
\begin{equation}
\label{Pi-expr}
\Pi_m \ = \
\frac{i}{2}\left(\pi^J\tilde\sigma_{m}\pi_J \right)\left(\pi^I\frac{\partial}{\partial\pi^I}- \rho^A\frac{\partial}{\partial\rho^A}\right)
+2i\left(\pi^I\tilde\sigma_{m}\rho^A \right)
\left(\!\pi_I\frac{\partial}{\partial\rho^A}\!\right)
\,.
\end{equation}
As a result, the square of $\Pi_m$ equals
\begin{equation}
\label{Pi-2}
C_4=\Pi^m\Pi_m = 2 \left(\epsilon^{\alpha\beta\gamma\delta} \pi_\alpha^K\pi_\beta{}_K \rho_\gamma^C\rho_\delta{}_C\right)
\epsilon^{IJ}\epsilon^{AB}
\left(\!\pi_I\frac{\partial}{\partial\rho^A}\!\right) \left(\!\pi_J\frac{\partial}{\partial\rho^B}\!\right)
\,.
\end{equation}

Now it is clear that the condition \p{C4-ir}: $C_4=-\mu^2$
is fulfilled on the infinite spin states if the
 following equations
for the infinite spin field $\Psi(\pi,\rho)$ hold:
\begin{equation}
\label{eq-2}
\left( \pi_\alpha{}_I\frac{\partial}{\partial\rho_\alpha^A} \ -\ \frac{i}{2}\,\epsilon_{IA} \right) \Psi=0 \,,
\end{equation}
\begin{equation}
\label{eq-1}
\Big( \epsilon^{\alpha\beta\gamma\delta} \pi_\alpha^K\pi_\beta{}_K \rho_\gamma^C\rho_\delta{}_C \ - \ \mu^2 \Big)\,
\Psi=0 \,.
\end{equation}

Equation \p{eq-2} leads to an important corollary due to the
presence of the tensor $\epsilon_{IA}$ in it. This tensor
is invariant only if the same $\mathrm{SU}(2)$
group acts on the indices $I$ and $A$
of the spinors $\pi_\alpha^I$ and $\rho_\alpha^A$, respectively.
Therefore, when we describe infinite spin
representations by means of
equations \p{eq-2} and \p{eq-1}, the
indices $I$ and $A$ in the spinors $\pi_\alpha^I$ and $\rho_\alpha^A$
refer to the same group $\mathrm{SU}(2)$ and these
indices can be covariantly
contracted to each other.
One can say that the whole automorphism group
$SU(2) \times SU(2)$ is reduced to the
diagonal subgroup $SU(2) \subset SU(2) \times SU(2)$.

Let us make some comments regarding the choice of the conditions \p{eq-2}, \p{eq-1}.
The operator \p{Pi-2} is the product of two commuting operators
$$
\Delta:=\epsilon^{\alpha\beta\gamma\delta} \pi_\alpha^K\pi_\beta{}_K \rho_\gamma^C\rho_\delta{}_C\,, \qquad
F:=2\epsilon^{IJ}\epsilon^{AB}
\left(\!\pi_I\frac{\partial}{\partial\rho^A}\!\right) \left(\!\pi_J\frac{\partial}{\partial\rho^B}\!\right),
$$
which can be diagonalized simultaneously.
Consequently, the fulfillment of the condition \p{C4-ir} is guaranteed by fixing
$\Delta\Psi=\mu^2\Psi$ \p{eq-1} and
\begin{equation}
\label{eq-2aa}
F\Psi=-\Psi \,.
\end{equation}
To satisfy the second order differential condition \p{eq-2aa}, we
require the fulfillment of stronger conditions \p{eq-2} of the
first order in the derivatives which are simpler than the condition
\p{eq-2aa}. In addition, equations \p{eq-2} are direct $6D$
generalization of the twistor equations for infinite spin fields in
four dimensions \cite{BFIR,BFI}. Although equations \p{eq-2} do
not cover all the solutions of \p{eq-2aa}, the choice of  \p{eq-2} allows us to
consider all possible infinite spin representations in six
dimensions, as will be seen below.\footnote{Note that
relations \p{eq-2}, \p{eq-1} are the sufficient conditions for the
irreducible infinite spin representation. We do not discuss here the
necessary conditions that do not affect the results of this paper.}

It remains to achieve the fulfillment of condition \p{C6-ir} for the
six-order Casimir operator on the infinite spin fields $\Psi(\pi,\rho)$.
Derivation of the expression for the six-order Casimir operator
\p{C6-Cas} in the representations \p{P-2} and \p{M-2} is technically
cumbersome. The main steps in this calculation are given in
Appendix\,C. As a result, we obtain that the action
of the six-order Casimir operator \p{C6-Cas}
on the two-twistor field $\Psi(\pi,\rho)$, satisfying
 conditions
\p{eq-2} and \p{eq-1}, gives
\begin{equation}\label{C6-Cas-ap-fin}
C_6\Psi=-\, \mu^2 \, J_aJ_a\,\Psi \,,
\end{equation}
where the operators $J_a$ $(a=1,2,3)$ have the form
\begin{equation}\label{Ta-def}
J_a \ := \ \frac12\,\pi_\alpha^I(\sigma_a)_I{}^J \frac{\partial}{\partial\pi_\alpha^J}
\ + \ \frac12\,\rho_\alpha^A (\sigma_a)_A{}^B \frac{\partial}{\partial\rho_\alpha^B}
\end{equation}
and $\sigma_a$ are the Pauli matrices. The operators $J_a$ generate
the $\mathfrak{su}(2)$ algebra
\begin{equation}\label{Ta-alg}
[J_a,J_b]= i\epsilon_{abc}J_c\,.
\end{equation}
Comparing condition \p{C6-Cas-ap-fin} with relation \p{C6-ir},
we find that when describing the irreducible infinite spin
representations, the twistor field $\Psi(\pi,\rho)$ must obey the
following condition:
\begin{equation}\label{eq-3}
J_aJ_a\,\Psi \ = \ s(s+1)\,\Psi \,,
\end{equation}
where the operators $J_a$ are defined in \p{Ta-def}. So in the
definition \p{C6-ir} of irreducible infinite spin representations,
the quantum number $s$  coincides with the spin, which labels the
representations of the diagonal $\mathrm{SU}(2)$ automorphism
subgroup. This subgroup is generated by the Lie algebra
$\mathfrak{su}(2)$ with the basis elements \p{Ta-def}.

Thus, we find a {\it full set} of equations on
the twistor field that describe the space of
massless infinite spin representations. Namely, the irreducible
infinite spin representations (with the quantum numbers $\mu$ and $s$)
of the Poincar\'e algebra under consideration with generators
\p{P-2}, \p{M-2} act in the space of the two-twistor fields
$\Psi(\pi,\rho)$ subject to equations \p{eq-2}, \p{eq-1} and
\p{eq-3}.

\setcounter{equation}0
\section{Twistorial infinite spin fields}

In this Section, we obtain a general solution to equations
\p{eq-2}, \p{eq-1} and \p{eq-3}.
 Recall, that according to equation  \p{eq-3},
the twistor field $\Psi$ possesses the $\mathrm{SU}(2)$-spin
equal to $s$.
The field of this type can only be described by means
of the completely symmetric $2s$ rank
spin-tensor field
\begin{equation}
\label{Psi-cov}
 \Psi_{I_1\ldots I_{2s}}(\pi,\rho)
 =\Psi_{(I_1\ldots I_{2s})}(\pi,\rho)\,,
\end{equation}
where $I_i$ are $\mathrm{SU}(2)$-indices,
or equivalently by the $(2s+1)$-component twistor field
\begin{equation}
\label{Psi-m}
\Psi_{\mathrm{m}}=\Psi_{\mathrm{m}}(\pi,\rho)\,,\qquad \mathrm{m}=-s,-s+1,\ldots s-1,s\,,
\end{equation}
which behaves under the action of the $\mathfrak{su}(2)$ generators \p{Ta-def} as follows:
\begin{equation}
\label{J-Psi-m}
J_3\Psi_{\mathrm{m}}=\mathrm{m}\Psi_{\mathrm{m}}\,,\qquad
J_\pm\Psi_{\mathrm{m}}=\sqrt{(s\mp\mathrm{m})(s\pm\mathrm{m}+1)}\,
\Psi_{\mathrm{m}\pm 1}\,,
\end{equation}
where $J_\pm=J_1\pm iJ_2$.
Equations \p{J-Psi-m} yield the eigenvalue of the $\mathfrak{su}(2)$ Casimir operator equal to the same value as in \p{eq-3}.

For us, it is more convenient to use the
description where the twistor field \p{Psi-m} is
represented by the completely symmetric spin-tensor
field \p{Psi-cov}.
Namely, the irreducible $6D$ infinite spin representation with the labels
$\mu$ and $s$ is described by the twistor field \p{Psi-cov}, which is
a totally symmetric $\mathrm{SU}(2)$ spin-tensor
of rank $2s$.
These fields by definition satisfy the equations
\begin{equation}
\label{J-Psi-I}
J_a\Psi_{I_1\ldots I_{2s}}=
 -\frac{1}{2}\, \sum_{\ell=1}^{2s} \,(\sigma_a)_{I_\ell}{}^K
 \Psi_{(I_1...I_{\ell-1} K I_{\ell+1}... I_{2s})}
 \equiv -\frac{2s+1}{2}\,(\sigma_a)_{(I_1}{}^K
 \Psi_{KI_2\ldots I_{2s})} \,,
\end{equation}
where $J_a$ were given in \p{Ta-def}
and $\sigma_a$ are the Pauli matrices.
Equations \p{J-Psi-I}
are the same as equations \p{J-Psi-m} rewritten for the
$\mathrm{SU}(2)$ spin-tensor fields $\Psi_{\mathrm{m}}$.
In addition to relations \p{J-Psi-I}, the twistor field \p{Psi-cov} also
obeys equations \p{eq-1},
\p{eq-2}, i.e. the following equations:
\begin{equation}
\label{eq-12-cov}
\mathrm{a)}\ \ \ \mathcal{M}\, \Psi_{I_1\ldots I_{2s}}=0 \,, \qquad\qquad
\mathrm{b)}\ \ \ \mathcal{F}_{KL}\, \Psi_{I_1\ldots I_{2s}}=0 \,,
\end{equation}
containing the operators
\begin{eqnarray}
\label{op-eq-1-cov}
\mathcal{M} &:=& \epsilon^{\alpha\beta\gamma\delta} \pi_\alpha^K\pi_\beta{}_K \rho_\gamma^C\rho_\delta{}_C \ - \ \mu^2  \,,
\\ [5pt]
\label{op-eq-2-cov}
\mathcal{F}_{IA} &:=& \pi_\alpha{}_I\frac{\partial}{\partial\rho_\alpha^A} \ -\ \frac{i}{2}\,\epsilon_{IA}  \,.
\end{eqnarray}
Linear equations \p{J-Psi-I}, \p{eq-12-cov}
are self-consistent since all nonzero commutators of
the operators $(J_a,\mathcal{F}_{IK}, \mathcal{M})$,
appearing in the definition of these equations, are
\begin{equation}
\label{J-F-com}
[J_a,J_b]= i\epsilon_{abc}J_c\, , \quad\quad
\left[J_a, \mathcal{F}_{IK}\right]=-\frac12\,(\sigma_a)_{I}{}^L \mathcal{F}_{(KL)}-\frac12\,(\sigma_a)_{K}{}^L \mathcal{F}_{(IL)}\, ,
\end{equation}
and this means that the algebra of these operators is closed.
Note that the second equation in \p{J-F-com}
is in agreement with  \p{J-Psi-I}.

Let us clarify the geometric sense of the obtained twistor equations \p{J-Psi-I}, \p{eq-12-cov}.
\begin{description}
\item[i)]
A natural way to solve equation (\ref{eq-12-cov}\,a) is to
introduce the delta function $\delta(\mathcal{M})$ as a factor in the
twistor field $\Psi_{I_1\ldots I_{2s}}(\pi,\rho)$. This means that
the twistor field $\Psi_{I_1\ldots I_{2s}}(\pi,\rho)$ lives on the
surface $\mathcal{M}=0$. However, this condition $\mathcal{M}=0$ is in
fact a condition that the determinant of the
$(4{\times}4)$ matrix
\begin{equation}
\label{g-su4}
g \ := \  \parallel\! g_\alpha{}^{\overline{\beta}} \!\parallel \  :=  \  \sqrt{2/\mu}\,\left. \Big( \pi_\alpha{}^I \right| \rho_\alpha{}^B \Big) \, ,
\end{equation}
is equal to unity, $\det g=1$. In the definition \p{g-su4}
of the matrix $g$, we use the notation
for the $(4{\times}2)$ blocks
$(g_\alpha{}^I): = (\pi_\alpha{}^I)$ and
$(g_\alpha{}^{2+B}):=(\rho_\alpha{}^B)$,
and the upper index $\overline{\beta}=1,2,3,4$
(the number of rows of the matrix $g$) is related to
the automorphism group $SU(2)$.
Moreover, the ``reality'' conditions in \p{pi} and \p{rho}
are written (with the help of the matrix $g$) in a concise form
\begin{equation}
\label{g-su4-1}
Bg= g^*{} \overline{B} \,, \quad\quad\qquad
\overline{B}_{\overline{\alpha}}{}^{\overline{\beta}} \equiv
B_\alpha{}^{\beta} \; ,
\end{equation}
where the matrix $B$ is given in \p{B-ap}
and the matrix $\overline{B}$ coincides with $B$
but acts in the space of the representation
of the automorphism group $SU(2)$.
Therefore, in view of the condition
$\det(g)=1$, the matrices \p{g-su4} belong to the $\mathrm{SU}^*(4)$ group where the matrix $B$
characterizes $\mathrm{SU}^*(4)$
involutive automorphism \cite{KuTow} (see also \cite{IR0},
Sect.\,3.3.2).
\item[ii)]
Condition (\ref{eq-12-cov}\,b) means that the following property
\begin{equation}
\label{sym-twf-1}
\Psi_{I_1\ldots I_{2s}}(\pi_\alpha^K,\rho_\alpha^B+\pi_\alpha^L\kappa_L{}^B) =
\exp\{i \kappa_L{}^L/2\}\Psi_{I_1\ldots I_{2s}}(\pi_\alpha^K,\rho_\alpha^B)
\end{equation}
is fulfilled for the twistor field where the matrix $\kappa_L{}^B$  obeys the reality condition of the form $(\kappa_L{}^B)^*=\epsilon^{LJ}\kappa_J{}^C\epsilon_{CB}$.
\item[iii)]
Equations \p{J-Psi-I} give the conditions for the $\mathrm{SU}(2)$ covariance of the twistor field:
\begin{equation}
\label{su2-twf-cov}
\Psi_{I_1\ldots I_{2s}}(\pi_\alpha^L u_L{}^K,\rho_\alpha^B u_B{}^A) =
u_{I_1}{}^{J_1} \ldots u_{I_{2s}}{}^{J_{2s}} \Psi_{J_1\ldots J_{2s}}(\pi_\alpha^K,\rho_\alpha^A) \,,
\end{equation}
where $u_I{}^K$ and $u_A{}^B$  are the same matrix belonging to the $\mathrm{SU}(2)$ group.
\end{description}

The transformations of the arguments of the twistor field in \p{sym-twf-1} and \p{su2-twf-cov}
are related to the right-hand transformations of the matrix $g$ \p{g-su4}
by the upper-triangular matrix $h$:
\begin{equation}
\label{su4-coset-tr}
g \to gh\,, \qquad h \ := \  \parallel\! h_{\overline{\alpha}}{}^{\overline{\beta}} \!\parallel \  :=  \
\left(
\begin{array}{cc}
u_I{}^J & \kappa_I{}^B \\
0 & u_A{}^B \\
\end{array}
\right)
\, ,
\end{equation}
where the diagonal blocks of $h$ are the same $\mathrm{SU}(2)$ matrix $u$, $uu^\dagger=1$, $\det u=1$,
and the $(2{\times}2)$ matrix $\kappa$ satisfies the condition $\kappa^*=\sigma_2\kappa\sigma_2$.
The matrices $h$ form the stability subgroup $\mathcal{K}\subset\mathrm{SU}^*(4)$.
This means that the twistor field
$\Psi_{I_1\ldots I_{2s}}(\pi,\rho)$ is a function on the coset $\mathrm{SU}^*(4)/\mathcal{K}$.
The points on this coset are parametrized by elements \p{g-su4} of the $\mathrm{SU}^*(4)$ group.
In addition, the notation ${\overline{\beta}}$ in  \p{g-su4} for the second index of the matrix $g_\alpha{}^{\overline{\beta}}$
differs from the first index  $\alpha$ and indicates that the bar-indices are related to the stability subgroup
$\mathcal{K}\subset\mathrm{SU}^*(4)$.

The solution of equations (\ref{eq-12-cov}\,a) and (\ref{eq-12-cov}\,b) can be represented
in the following form.

Similar to the four-dimensional case \cite{BFIR,BFI}, equation
(\ref{eq-12-cov}\,a) is solved by taking the delta-function
$\delta(\epsilon^{\alpha\beta\gamma\delta} \pi_\alpha^K\pi_\beta{}_K
\rho_\gamma^C\rho_\delta{}_C - \mu^2)$ as a factor in the twistor
field, while equation (\ref{eq-12-cov}\,b) is greatly simplified
by extracting the exponent $\exp(iu_0{}^I{}_I/v_0)$, where
$v_0=\left(\pi^I\tilde\sigma_{0}\pi_I \right)$,
$u_0{}^{\!I}{}_{\!I}=\left(\pi^I\tilde\sigma_{0}\rho_I \right)$ (see
\p{vectors}). Thus, we consider the field
\begin{equation}
\label{Psi-cov-sol}
\Psi_{I_1\ldots I_{2s}}(\pi,\rho)=
\delta(\epsilon^{\alpha\beta\gamma\delta} \pi_\alpha^K\pi_\beta{}_K \rho_\gamma^C\rho_\delta{}_C - \mu^2)
\,e^{\displaystyle iu_0{}^{\!I}{}_{\!I}/v_0}\,
\tilde\Psi_{I_1\ldots I_{2s}}(\pi,\rho)\,,
\end{equation}
where the field $\tilde\Psi_{I_1\ldots I_{2s}}(\pi,\rho)$ satisfies the equations
\begin{equation}
\label{J-tilde-Psi-col}
J_a\tilde\Psi_{I_1\ldots I_{2s}}=-\frac{2s+1}{2}\,(\sigma_a)_{(I_1}{}^K\tilde\Psi_{KI_2\ldots I_{2s})}\,,
\end{equation}
\begin{equation}
\label{eq-12-cov-sol}
\pi_\alpha{}_I\frac{\partial}{\partial\rho_\alpha^A} \, \tilde\Psi_{I_1\ldots I_{2s}}(\pi,\rho)=0 \,.
\end{equation}
As a result, the field $\tilde\Psi_{I_1\ldots I_{2s}}(\pi,\rho)$ is
represented by an infinite series where each term has the
$\mathrm{SU}(2)$  spin $s$ and its indices are represented by the
tensor product of the spinor indices of $\pi_{\alpha}^I$ and
$\rho_{\alpha}^A$ and their products. For example, in the case
$s{=}1/2$, the field $\tilde\Psi_{I_1\ldots I_{2s}}(\pi,\rho)$ has
infinite expansion with the terms
\begin{equation}
\label{tildePsi-1/2}
\tilde\Psi_{I}(\pi,\rho)=\pi_{\alpha}{}_I \psi^\alpha+
\epsilon^{\alpha\beta\gamma\delta}\pi_{\beta}^K\pi_{\gamma}{}_K\rho_{\delta}{}_I\varphi_\alpha+\
\pi_{\alpha}{}_I \epsilon^{\beta_1\beta_2\gamma_1\gamma_2}\pi_{\gamma_1}^K\pi_{\gamma_2}{}_K \psi^\alpha_{\beta_1\beta_2}
+\ldots
\,,
\end{equation}
where $\psi^\alpha$, $\varphi_\alpha$,
$\psi^\alpha_{\beta_1\beta_2}$, $\ldots$ describe the infinite
degrees of freedom of the infinite spin representation.

Summing up this section,
we conclude that twistor fields \p{Psi-cov-sol}, \p{J-tilde-Psi-col},
\p{eq-12-cov-sol}, forming the space of irreducible massless infinite spin
representations for the Poincar\'e group in six dimensions, were constructed.

\setcounter{equation}0
\section{Field twistor transform and space-time infinite spin fields}

An important task in the description of infinite spin representation is
finding the field twistor transform that links the twistor field
formulation with the space-time one  (see some discussion e.g.
in \cite{BekMou,BekBoul}).

One of the possible $6D$ space-time descriptions of the
twistor transform uses the additional spinorial coordinates. Namely,
using the twistor fields $\Psi_{I_1\ldots I_{2s}}(\pi,\rho)$
(\ref{Psi-cov}), we can construct the fields
\begin{equation}
\label{tw-f-scalar}
\pi_{\alpha_1}^{I_1}\ldots \pi_{\alpha_{2s}}^{I_{2s}}\,\Psi_{I_1\ldots I_{2s}}(\pi,\rho) \,,
\end{equation}
which are $\mathrm{SU}(2)$ scalars:
\begin{equation}
\label{cond-tw-f-scalar}
J_a\left(\pi_{\alpha_1}^{I_1}\ldots \pi_{\alpha_{2s}}^{I_{2s}}\,\Psi_{I_1\ldots I_{2s}} \right)=0 \,.
\end{equation}
Then, performing the following integral transform
\begin{equation}
\label{f-trans}
\Phi_{\alpha_1\ldots \alpha_{2s}}(x,\rho) =
\int \mu(\pi)\, {e}^{\displaystyle ix^m p_m} \,\pi_{\alpha_1}^{I_1}\ldots \pi_{\alpha_{2s}}^{I_{2s}}\,\Psi_{I_1\ldots I_{2s}}(\pi,\rho) \,,
\end{equation}
where
\begin{equation}
\label{mu-mes}
\mu(\pi) := \mu_{11}\wedge\mu_{22}\wedge\mu_{33}\wedge\mu_{44} \,,\qquad
\mu_{\alpha\beta}:=\frac12\,d\pi^I_{\alpha}\wedge d\pi_{\beta}{}_I
\end{equation}
is a real integration measure in ``$\pi$-space'', which can also be written in the form
$
\mu(\pi)\sim \epsilon^{\alpha\beta\gamma\delta}\epsilon^{\mu\nu\lambda\rho}
\mu_{\alpha\mu}\wedge\mu_{\beta\nu}\wedge\mu_{\gamma\lambda}\wedge\mu_{\delta\rho}
$\,,
and $p_m=\pi_\alpha^I (\tilde\sigma_{m})^{\alpha\beta}\pi_\beta{}_I$
as in \p{P-2}, we obtain a completely symmetric space-time field
$\Phi_{\alpha_1\ldots \alpha_{2s}}(x,\rho)$, which depends on the
space-time coordinates $x^m$ and additional spinor variables
$\rho_\alpha^I$.

The field \p{f-trans}
automatically satisfies the Dirac equation
\begin{equation}
\label{f-Dir}
i\frac{\partial}{\partial x^m}\,(\tilde\sigma^m)^{\beta\alpha_1}
\Phi_{\alpha_1\ldots \alpha_{2s}}(x,\rho) = 0  \; ,
\end{equation}
that yields the Klein-Gordon equation
\begin{equation}
\label{f-KG}
\frac{\partial}{\partial x^m}\frac{\partial}{\partial x_m}\,
\Phi_{\alpha_1\ldots \alpha_{2s}}(x,\rho) = 0 \; .
\end{equation}
for all values of the spin $s{>}\,0$, while for $s{=}\,0$
eq.\,\p{f-KG} is checked directly. For all $s$ equations \p{f-Dir} and
\p{f-KG} follow from the identity
$\epsilon^{\alpha\beta\gamma\delta}\pi_\alpha^I
\pi_\beta^J\pi_\gamma^K\equiv 0$.

Two other equations of motion for the space-time field \p{f-trans} are obtained from equations (\ref{eq-12-cov}) for the twistor field
$\Psi_{I_1\ldots I_{2s}}(\pi,\rho)$.
Namely, the twistor condition (\ref{eq-12-cov}\,a)
is equivalent to the equation
\begin{equation}
\label{f-is-1}
\left(i\frac{\partial}{\partial x^m}\ \rho_\beta^K(\tilde\sigma^m)^{\beta\gamma}\rho_\gamma{}_K \ + \ 2\mu^2\right) \Phi_{\alpha_1\ldots \alpha_{2s}} =
0\,,
\end{equation}
while the equation
\begin{equation}
\label{f-is-2}
\left(i\frac{\partial}{\partial x^m}\ \frac{\partial }{\partial \rho_\beta^K}(\sigma^m)_{\beta\gamma}
\frac{\partial }{\partial \rho_\gamma{}_K}\ - \ 2\right)\Phi_{\alpha_1\ldots \alpha_{2s}} =0
\end{equation}
is a consequence of the twistor condition (\ref{eq-12-cov}\,b).

In addition, the space-time field \p{f-trans} obeys the equations
\begin{equation}
\label{f-is-3}
\rho_\beta^I \; (\sigma_a)_{_I}{}^{_K} \;
 \frac{\partial }{\partial \rho_\beta^K}
\ \Phi_{\alpha_1\ldots \alpha_{2s}} =0\,.
\end{equation}
To deduce equation \p{f-is-3},
we use solution \p{Psi-cov-sol} for the twistor field, which
is included in the integrand in the right-hand side of
\p{f-trans}.
Then the left-hand side of equations \p{f-is-3}
takes the form
\begin{eqnarray}
&&
\left(\rho_\nu \sigma_a
 %\frac{\partial }{\partial \rho_\nu}
  \partial_{\rho_\nu}
\right)\Phi_{\alpha_1\ldots \alpha_{2s}} =  \int \mu(\pi)\, {e}^{\displaystyle ix^m p_m}
\left[ \left(\rho_\nu \sigma_a
  \partial_{\rho_\nu} \right)
e^{\displaystyle iu_0{}^{\!I}{}_{\!I}/v_0}\right]
\delta(\epsilon^{\alpha\beta\gamma\delta} (\pi_\alpha\pi_\beta) (\rho_\gamma\rho_\delta)- \mu^2)
\,\tilde\Psi_{\alpha_1\ldots \alpha_{2s}}
\nonumber \\
&&+\int \mu(\pi)\, {e}^{\displaystyle ix^m p_m}
\, e^{\displaystyle iu_0{}^{\!I}{}_{\!I}/v_0}
\left[\left(\rho_\nu \sigma_a
  \partial_{\rho_\nu} \right)
\delta\bigl(\epsilon^{\alpha\beta\gamma\delta}
(\pi_\alpha\pi_\beta) (\rho_\gamma\rho_\delta) - \mu^2
\bigr)
\,\tilde\Psi_{\alpha_1\ldots \alpha_{2s}}\right] ,
\label{ap-1}
\end{eqnarray}
where we
use the concise notation
$\tilde\Psi_{\alpha_1\ldots \alpha_{2s}}:=
\pi_{\alpha_1}^{I_1}\ldots \pi_{\alpha_{2s}}^{I_{2s}}\tilde\Psi_{I_1\ldots I_{2s}}$,
$\displaystyle \partial_{\rho_\beta^K} :=
\frac{\partial }{\partial \rho_\beta^K}$ and
omit the contracted $SU(2)$ spinor indices
$(\pi_\alpha\pi_\beta) := \pi_\alpha^K\pi_\beta{}_K$,
$(\rho_\gamma\rho_\delta) := \rho_\gamma^C\rho_\delta{}_C$,
$\displaystyle \rho_\beta^I (\sigma_a)_{_I}{}^{_K} \frac{\partial }{\partial \rho_\beta^K}:= (\rho_\beta \sigma_a \partial_{\rho_\beta})$.
Taking into account the identity $J_a\left(u_0{}^{\!I}{}_{\!I}/v_0\right)=0$,
we represent the first term in the right hand side of
\p{ap-1} as
\begin{eqnarray}
\label{ap-2}
&&-\int \mu(\pi)\, {e}^{\displaystyle ix^m p_m}
\left[ \left(\pi_\nu \sigma_a \partial_{\pi_\nu}\right)
e^{\displaystyle iu_0{}^{\!I}{}_{\!I}/v_0}\right]
\delta\bigl(\epsilon^{\alpha\beta\gamma\delta}
(\pi_\alpha\pi_\beta) (\rho_\gamma\rho_\delta) - \mu^2\bigr)
\,\tilde\Psi_{\alpha_1\ldots \alpha_{2s}} \; .
\nonumber
\end{eqnarray}
Finally, after integration by parts in this expression,
equality \p{ap-1} takes the form
\begin{equation}
\left(\rho_\nu \sigma_a \partial_{\rho}{}^{\nu}\right)
\Phi_{\alpha_1\ldots \alpha_{2s}}  =
2\!\int \!\! \mu(\pi) \, {e}^{\displaystyle ix^m p_m}
e^{\displaystyle iu_0{}^{\!I}{}_{\!I}/v_0}
\Bigl[J_a\,
\delta\Bigl(\epsilon^{\alpha\beta\gamma\delta}
(\pi_\alpha\pi_\beta) (\rho_\gamma\rho_\delta) - \mu^2\Bigr)
\,\tilde\Psi_{\alpha_1\ldots \alpha_{2s}}\Bigr].
\label{ap-3}
\end{equation}
However, due to relations \p{cond-tw-f-scalar} and
$J_a\,\delta\Bigl(\epsilon^{\alpha\beta\gamma\delta} (\pi_\alpha\pi_\beta) (\rho_\gamma\rho_\delta) - \mu^2\Bigr)=0$,
the right-hand side of equality \p{ap-3} is equal to zero; therefore, we obtain \p{f-is-3}.

Thus, we have proposed
the space-time formulation (with additional spinor variables
$\rho^I$)
of the $6D$ infinite spin fields
which is a generalization of our formulation of the $4D$
infinite spin fields \cite{BFIR,BFI,BuchIFKr}.
In the framework of this formulation,  we obtained the
equations of motion \p{f-Dir} -- \p{f-is-3}
for the $6D$ infinite spin fields \p{f-trans}.
Equations \p{f-Dir} -- \p{f-is-2} are
6D analogs of the corresponding equations in
the four-dimensional case. The new $\mathrm{SU}(2)$ conditions \p{f-is-3} replace the $\mathrm{U}(1)$ condition in the $4D$ case.
Also, in contrast to the $4D$ case, the $6D$ infinite spin fields in the space-time formulation  have external spinor indices
$(\alpha_1,...,\alpha_{2s})$,
the number of which is determined by the quantum number $s$.

\setcounter{equation}0
\section{Summary and outlook}

Let us summarize the results. We have studied the twistor field
realization of the Poincar\'{e} group massless irreducible
representations of infinite spin in the $6D$ Minkowski space. It
was shown that infinite spin representations are realized on the
twistor fields living in the two-twistor space, which may be
interesting in the context of the higher spin field theory. The group
generators and the corresponding Casimir operators were derived in
the considered two-twistor realization. As result, we found a full set
of equations of motion \p{J-Psi-I}, \p{eq-12-cov} for the infinite spin
twistor field \p{Psi-cov} that is the totally symmetric
$\mathrm{SU}(2)$ tensor of rank $2s$. Thus, we believe that the
irreducible massless representations under consideration
are completely described in terms of twistor fields. In addition, using
the field twistor transform, we determine space-time fields that
describe infinite spin representations in the formulation with an
additional spinor coordinate.

Note that the spinor $\rho_{\alpha}^A$ \p{rho}, corresponding to the
introduced second twistor, has the same $\mathrm{SU}^*(4)$ chirality
as the spinor  $\pi_{\alpha}^I$ \p{pi} of the first twistor. In
principle, we can introduce the second twistor of different chirality.
As one can see, this situation arises after the transition from the spinor
variable $\rho_{\alpha}^A$ \p{rho} to the spinor variable
$\eta^{\alpha}_A$ \p{d-rho} with the help of a simple Fourier transform
with respect to these variables. We are going to give a complete
answer to this question in subsequent papers.

As a continuation of this research, it would be interesting to
construct one-dimensional
classical and quantum mechanics of an infinite spin particle in both
space-time and twistor formulations, similar to our works
\cite{BFIR,BFI} for the $4D$ case. This will allow us to derive
the twistor transforms that provide correspondence of the phase
variables in space-time with twistor patterns. In addition,
analysis of the quantum twistor wave function will make it possible
to analyze the spin expansion of the twistor field.
In the forthcoming papers, we plan also to study the relation of our twistor model to
another $6D$ space-time  formulation with additional vector
variables, which was proposed in \cite{BekMou,BekBoul} (see also
\cite{BFIP}).

\section*{Acknowledgments}
The work of ILB and SAF is supported by the Russian Science Foundation, project No 21-12-00129.
API was partially supported by the Ministry of Education of the Russian Federation, project FEWF-2020-0003.

%\newpage

\section*{Appendix\,A: \ $6D$ spinor notation }
\def\theequation{A.\arabic{equation}}
\setcounter{equation}0

\quad\,
In this Appendix we present the $6D$ spinor notation used in this paper.

In the Weyl representation, the $(8{\times}8)$ Dirac $\gamma$-matrices have the form
\begin{equation}
\label{gamma-matr}
\gamma^m=
\left(
\begin{array}{cc}
0 & (\sigma^m)_{\alpha\dot\beta} \\
(\tilde\sigma^m)^{\dot\alpha\beta} & 0 \\
\end{array}
\right)\,,
\qquad \{\gamma^m,\gamma^n\}=2\,\eta^{mn}\,,
\end{equation}
where the Weyl indices run four values: $\alpha,\dot\alpha=1\dots,4$.
We take the following representation \cite{KuTow,IR}:
$$
\sigma^m=(\sigma^0,\sigma^a)\,,\qquad \tilde\sigma^m=(\sigma^0,-\sigma^a)\,,\qquad a=1,\ldots,5\,,
$$
where $\sigma^0=1_4$, $\sigma^1=\tau_1\otimes 1_2$, $\sigma^{\hat a}=\tau_2\otimes \tau_{\hat a-1}$ at $\hat a=2,3,4$,
$\sigma^5=\tau_3\otimes 1_2$, and $\tau_{1,2,3}$ are the Pauli matrices.

The matrices $\mathcal{B}$ and $\mathcal{C}$ defining $(\gamma^m)^*=-\mathcal{B}\gamma^m \mathcal{B}^{-1}$ and
$(\gamma^m)^T=-\mathcal{C}\gamma^m \mathcal{C}^{-1}$
have the form
$$
\mathcal{B}=\left(
\begin{array}{cc}
B & 0 \\
0 & B^T \\
\end{array}
\right)\,,
\quad
\mathcal{C}=
\left(
\begin{array}{cc}
0 & B^T\\
B & 0 \\
\end{array}
\right)\,,
$$
where
\begin{equation}
\label{B-ap}
B=\| B_{\dot\alpha}{}^{\beta}\|=1_2\otimes i\tau_{2}=
\left(
\begin{array}{cc}
i\tau_2 & 0 \\
0 & i\tau_2 \\
\end{array}
\right)\,.
\end{equation}
Using quantities \p{B-ap} we can construct the matrices
\begin{equation}
\label{sigma-m}
(\sigma^{m})_{\alpha\beta}=(\sigma^{m})_{\alpha\dot\gamma}(B^{-1})_\beta{}^{\dot\gamma}\,,\qquad
(\tilde\sigma^{m})^{\alpha\beta}=B_{\dot\gamma}{}^\alpha(\tilde\sigma^{m})^{\dot\gamma\beta}\,,
\end{equation}
having only undotted indices. These matrices are antisymmetric,
$(\sigma^{m})_{\alpha\beta}=-(\sigma^{m})_{\beta\alpha}$,
$(\tilde\sigma^{m})^{\alpha\beta}=-(\tilde\sigma^{m})^{\beta\alpha}$
and satisfy the relations
\begin{equation}
\label{sigma-eq}
(\sigma^{m})_{\alpha\gamma}(\tilde\sigma^{n})^{\gamma\beta}+(\sigma^{n})_{\alpha\gamma}(\tilde\sigma^{m})^{\gamma\beta}
=2\,\eta^{mn}\delta^\beta_\alpha\,,
\end{equation}
\begin{equation}
\label{sigma-eq1}
\sigma^m_{\alpha\beta}\tilde\sigma_n^{\alpha\beta}=-4\,\delta^m_n\,,\qquad
(\sigma^{m})_{\alpha\beta}(\tilde\sigma_{m})^{\gamma\delta}=
-4\,\delta^\gamma_{[\alpha}\delta^\delta_{\beta]}\, .
\end{equation}
We use the convention for the antisymmetrization: $X_{[\alpha}Y_{\beta]}:=\frac12\,(X_{\alpha}Y_{\beta}-X_{\beta}Y_{\alpha})$.
Moreover, the matrices \p{sigma-m} are expressed through each other by means of the equations
\begin{equation}
\label{sigma-eq2}
(\sigma^{m})_{\alpha\beta}=-\frac12\,\varepsilon_{\alpha\beta\gamma\delta}(\tilde\sigma^{m})^{\gamma\delta}\, ,\qquad
(\tilde\sigma^{m})^{\alpha\beta}=-\frac12\,\varepsilon^{\alpha\beta\gamma\delta}(\sigma^{m})_{\gamma\delta}\, ,
\end{equation}
where $\varepsilon_{\alpha\beta\gamma\delta}$ and
$\varepsilon^{\alpha\beta\gamma\delta}$ are the totally
antisymmetric tensors with components
$\varepsilon_{1234}=\varepsilon^{1234}=1$. Therefore, the following
relations are fulfilled
\begin{equation}
\label{sigma-eq3}
(\sigma^{m})_{\alpha\beta}(\sigma_{m})_{\gamma\delta}= 2\,\varepsilon_{\alpha\beta\gamma\delta}\, ,\qquad
(\tilde\sigma^{m})^{\alpha\beta}(\tilde\sigma_{m})^{\gamma\delta}=2\,\varepsilon^{\alpha\beta\gamma\delta}\, .
\end{equation}

Transformations of the matrices \p{sigma-m} under complex
conjugation look like:
\begin{equation}
\label{sigma-m-cc}
[(\sigma^{m})_{\alpha\beta}]^*=B_{\dot\alpha}{}^\gamma B_{\dot\beta}{}^\delta (\sigma^{m})_{\gamma\delta}\,,\qquad
[(\tilde\sigma^{m})^{\alpha\beta}]^*=(\tilde\sigma^{m})^{\gamma\delta}(B^{-1})_\gamma{}^{\dot\alpha}(B^{-1})_\delta{}^{\dot\beta}\,.
\end{equation}

The $\sigma$-matrices with two vector indices are defined by
\begin{equation}
\label{sigma-mn}
(\sigma_{mn})_{\alpha}{}^{\beta}=\frac14\,(\sigma_m \tilde\sigma_{n}-\sigma_n \tilde\sigma_{m})_{\alpha}{}^{\beta}\,,\qquad
(\tilde\sigma_{mn})^{\alpha}{}_{\beta}=
-(\sigma_{mn})_{\beta}{}^{\alpha}=\frac14\,(\tilde\sigma_m \sigma_{n}-\tilde\sigma_n \sigma_{m})^{\alpha}{}_{\beta}\,.
\end{equation}
They satisfy the identities
\begin{equation}
\label{eps-sigma-mn}
\begin{array}{rcl}
[\sigma_{mn},\sigma_{kl}]&=&\eta_{nk}\sigma_{ml}-\eta_{nl}\sigma_{mk}+\eta_{mk}\sigma_{ln}-\eta_{ml}\sigma_{kn}\,,\\ [6pt]
[\tilde\sigma_{mn},\tilde\sigma_{kl}]&=&\eta_{nk}\tilde\sigma_{ml}-\eta_{nl}\tilde\sigma_{mk}+\eta_{mk}\tilde\sigma_{ln}-\eta_{ml}\tilde\sigma_{kn}\,.
\end{array}
\end{equation}
There are also the following equalities:
\begin{equation}
\label{sigma-eq1a}
(\tilde\sigma^{mn})^{\alpha}{}_{\beta}(\tilde\sigma_{mn})^{\gamma}{}_{\delta}=
\frac12\,\delta^{\alpha}_{\beta}\delta^{\gamma}_{\delta}-2\delta^{\alpha}_{\delta}\delta^{\gamma}_{\beta}\,,
\end{equation}
\begin{equation}
\label{sigma-eq2a}
(\tilde\sigma^{n})^{\alpha\beta}(\tilde\sigma_{nm})^{\gamma}{}_{\delta}=
\frac12\,(\tilde\sigma_{m})^{\alpha\beta}\delta^{\gamma}_{\delta}+2\delta^{[\alpha}_{\delta}(\tilde\sigma_{m})^{\beta]\gamma},\quad
(\sigma^{n})_{\alpha\beta}(\tilde\sigma_{nm})^{\gamma}{}_{\delta}=
-\frac12\,(\sigma_{m})_{\alpha\beta}\delta^{\gamma}_{\delta}-2\delta_{[\alpha}^{\gamma}(\sigma_{m})_{\beta]\delta}\,.
\end{equation}

The eight-component Dirac spinor $\Psi$ and its Dirac conjugate one $\bar\Psi=\Psi^\dagger\gamma_0$
are represented by two four-component Weyl spinors
\begin{equation}
\label{Dir-sp}
\Psi=\left(
\begin{array}{c}
\psi_\alpha \\ [5pt]
\chi^{\dot\alpha} \\
\end{array}
\right)
,\qquad
\bar\Psi=\left(\bar\chi^\alpha ,\bar\psi_{\dot\alpha} \right),\qquad\quad \bar\psi_{\dot\alpha}=(\psi_\alpha)^*\,,\quad
\bar\chi^{\alpha}=(\chi^{\dot\alpha})^*
\,.
\end{equation}
Opposite to the $D=1+3$ case, in the $D=1+5$ Minkowski case the spinors $\psi_\alpha$ and $\bar\psi_{\dot\alpha}$ realize
the same spinor representation.
Therefore, the spinorial components
$\psi_{\alpha}^{1}:=\psi_\alpha$, $\psi_{\alpha}^{2}:=(B^{-1})_\alpha{}^{\dot\beta}\bar\psi_{\dot\beta}$
form the $\mathrm{SU}(2)$ Majorana-Weyl spinor possessing the $\mathrm{SU}(2)$-reality ($ \epsilon_{12}=\epsilon^{21}=1$):
\begin{equation}
\label{su2MW-sp-cc}
(\psi_{\alpha}^{I})^* = \epsilon_{IJ}B_{\dot\alpha}{}^\beta \psi_{\beta}^{J}\,.
\end{equation}
Due to \p{sigma-m-cc} and \p{su2MW-sp-cc}, we find that the vector $\psi^I\tilde\sigma^{m}\psi_{I}$ is real
when spinor $\psi_{\alpha I}$ is even.

By analogy, defining
$\chi^{\alpha 1}:=\bar\chi^{\alpha}$, $\chi^{\alpha 2}:=\chi^{\dot\beta}B_{\dot\beta}{}^\alpha$,
we obtain the $\mathrm{SU}(2)$ Majorana-Weyl spinor with the $\mathrm{SU}(2)$-reality
\begin{equation}
\label{su2MW-sp-cc-1}
(\chi^{\alpha I})^* = \epsilon_{IJ}  \chi^{\beta J}(B^{-1})_\beta{}^{\dot\alpha}\,,
\end{equation}
and, therefore, the vector $\chi^{I}(\sigma^{m})\chi_{I}$ is real
at even spinor $\chi^{\alpha}_{I}$.

\section*{Appendix\,B: \ Twistor realization of $D{=}\,1{+}\,5$ conformal algebra}
\def\theequation{B.\arabic{equation}}
\setcounter{equation}0

Here we show that 28 quantities $X_{[ab]}$ (defined in \p{6-conf}) form the $\mathfrak{so}(2,6)$ algebra
with respect to the Poisson brackets \p{PB-tw}.

Generators\p{6-conf} are represented by the following set of quantities
\begin{equation}
\label{6-conf-1}
P_{\alpha\beta}:=\pi_\alpha^I \pi_\beta^J \epsilon_{IJ}\,,\qquad
K^{\alpha\beta}:=\omega^\alpha_I \omega^\beta_J \epsilon^{IJ}\,,\qquad
M_{\alpha}^{\beta}:=\pi_\alpha^I \omega^\beta_I\,,
\end{equation}
that have $D{=}\,6$ Weyl spinor indices.
Due to the $\mathrm{SU}(2)$ reality conditions \p{pi} and \p{omega},
 the set of the generators \p{6-conf-1}
goes into itself under complex conjugation.

Using the twistor Poisson brackets \p{PB-tw}, we obtain that the generators
\p{6-conf-1} obey the algebra
\begin{equation}
\label{PB-tw-1}
\begin{array}{c}
\left\{ P_{\alpha\beta}\,, P_{\gamma\delta} \right\}_{PB} = 0 \,, \qquad
\left\{ K^{\alpha\beta}\,, K^{\gamma\delta} \right\}_{PB} = 0 \,, \\ [6pt]
\left\{ P_{\alpha\beta}\,, K^{\gamma\delta} \right\}_{PB} = -\delta_\alpha^\gamma M_\beta^\delta +\delta_\alpha^\delta M_\beta^\gamma
+\delta_\beta^\gamma M_\alpha^\delta -\delta_\beta^\delta M_\alpha^\gamma \,,  \\ [6pt]
\left\{ M_{\alpha}^{\beta}\,, P_{\gamma\delta} \right\}_{PB} = -\delta_\gamma^\beta P_{\alpha\delta}-\delta_\delta^\beta P_{\gamma\alpha} \,, \qquad
\left\{ M_{\alpha}^{\beta}\,, K^{\gamma\delta} \right\}_{PB} = \delta_\alpha^\gamma K^{\beta\delta} + \delta_\alpha^\delta K^{\gamma\beta}\,,
\\ [6pt]
\left\{ M_{\alpha}^{\beta}\,, M_{\gamma}^{\delta} \right\}_{PB} = \delta_\alpha^\delta M_{\gamma}^{\beta} - \delta_\gamma^\beta M_{\alpha}^{\delta}\,.
\end{array}
\end{equation}
After passing from \p{6-conf-1} to pure real generators with vectorial indices
\begin{equation}
\label{6-conf-2}
P_m:=P_{\alpha\beta}(\tilde\sigma_m)^{\alpha\beta}\,,\quad
K_m:=\frac18\,K^{\alpha\beta}(\sigma_m)_{\alpha\beta}\,,\quad
M_{mn}:=-M_{\alpha}^{\beta}(\sigma_{mn})_{\beta}{}^{\alpha}\,,\quad
D:=-\frac12\,M_{\alpha}^{\alpha}\,,
\end{equation}
the algebra \p{PB-tw-1} takes the standard form of the $D{=}\,1{+}\,5$ conformal algebra:
\begin{equation}
\label{PB-tw-2}
\begin{array}{c}
\left\{ P_{m}\,, P_{n} \right\}_{PB} = 0 \,, \qquad
\left\{ K_{m}\,, K_{n} \right\}_{PB} = 0 \,, \\ [6pt]
\left\{ P_{m}\,, K_{n} \right\}_{PB} = 2\left( M_{mn} -\eta_{mn}D\right)\,,  \\ [6pt]
\left\{ M_{mn}\,, P_{k} \right\}_{PB} = \eta_{mk} P_{n} - \eta_{nk} P_{m} \,, \qquad
\left\{ M_{mn}\,, K_{k} \right\}_{PB} = \eta_{mk} K_{n} - \eta_{nk} K_{m}\,,
\\ [6pt]
\left\{ D\,, P_{m} \right\}_{PB} = P_{m} \,, \qquad
\left\{ D\,, K_{m} \right\}_{PB} = -K_{m}\,, \qquad
\left\{ D\,, M_{mn} \right\}_{PB} = 0\,,
\\ [6pt]
\left\{ M_{mn}\,, M_{kl} \right\}_{PB} = \eta_{mk} M_{nl} +\eta_{nl} M_{mk}- \eta_{ml} M_{nk} -\eta_{nk} M_{ml}\,.
\end{array}
\end{equation}
After the transition from the generators $M_{mn}$, $P_m$, $K_m$, $D$ to their linear combinations
$$
L_{mn}=M_{mn}\,,\quad L_{6m}=\frac12\left(P_m-K_m\right),\quad L_{7m}=\frac12\left(P_m+K_m\right),
\quad L_{67}=-D\,,
$$
algebra \p{PB-tw-2} takes the standard form for the $\mathfrak{so}(2,6)$ algebra
\begin{equation}
\label{so-26}
\left\{ L_{\mathcal{MN}}\,, L_{\mathcal{KL}} \right\}_{PB} = \eta_{\mathcal{MK}} L_{\mathcal{NL}} +\eta_{\mathcal{NL}} L_{\mathcal{MK}}-
\eta_{\mathcal{ML}} L_{\mathcal{NK}} -\eta_{\mathcal{NK}} L_{\mathcal{ML}} \,,
\end{equation}
where $\mathcal{M}=(m,6,7)$, $\eta_{66}=-\eta_{77}=-1$, $\eta_{6m}=\eta_{7m}=\eta_{67}=0$.

\section*{Appendix\,C: \ Calculation of $C_6$}
\def\theequation{C.\arabic{equation}}
\setcounter{equation}0

Using \p{M-2}, \p{Pi-expr} we obtain
\begin{equation}
\label{Pi-M-expr}
\Pi^n M_{nm} =
v_m A +
w_m B +
u^{IA}_m C_{IA}
\,,
\end{equation}
where
\begin{eqnarray}
\label{A-expr}
A &=&
-\frac{1}{4} \left(\pi^I\frac{\partial}{\partial\pi^I}- \rho^A\frac{\partial}{\partial\rho^A}\right)^2
- \left(\!\rho^A\frac{\partial}{\partial\pi^I}\!\right)
\left(\!\pi^I\frac{\partial}{\partial\rho^A}\!\right) , \\
\label{B-expr}
B &=&
-\epsilon^{IJ}\epsilon^{AB}
\left(\!\pi_I\frac{\partial}{\partial\rho^A}\!\right)
\left(\!\pi_J\frac{\partial}{\partial\rho^B}\!\right), \\
\label{C-expr}
C_{IA} &=&
2 \left(\rho^B\frac{\partial}{\partial\rho^B}+1\right)\left(\!\pi_I\frac{\partial}{\partial\rho^A}\!\right)
- 2\left(\!\pi^J\frac{\partial}{\partial\pi^I}\!\right)\left(\!\pi_J\frac{\partial}{\partial\rho^A}\!\right)
- 2\left(\!\rho^B\frac{\partial}{\partial\rho^A}\!\right)\left(\!\pi_I\frac{\partial}{\partial\rho^B}\!\right)
\end{eqnarray}
and vector variables $v_m$, $w_m$, $u^{IA}_m$ are defined in \p{vectors}.
The operators \p{A-expr}, \p{B-expr}, \p{C-expr} have the following commutators with these vector variables:
\begin{equation}\label{A-com}
\begin{array}{rcl}
\left[ A, v_m \right] & = &  \displaystyle
-v_m\left( \pi^J\frac{\partial}{\partial\pi^J} -
\rho^B\frac{\partial}{\partial\rho^B} +1\right)-2\,
u^{IA}_m\left( \pi_I\frac{\partial}{\partial\rho^A} \right) ,\\ [9pt]
\left[ A, w_m \right] & = &  \displaystyle
w_m\left( \pi^J\frac{\partial}{\partial\pi^J} -
\rho^B\frac{\partial}{\partial\rho^B} -5\right)+2\,
u^{IA}_m\left( \rho_A\frac{\partial}{\partial\pi^I} \right) ,\\ [9pt]
\left[ A, u^{IA}_m \right] & = & \displaystyle
-u^{IA}_m
+ \frac12\, v_m\,\epsilon^{IJ} \left( \rho^A\frac{\partial}{\partial\pi^J} \right)-
\frac12\, w_m\,\epsilon^{AB} \left( \pi^I\frac{\partial}{\partial\rho^B} \right) ,
\end{array}
\end{equation}
\begin{equation}\label{B-com}
\left[ B, v_m \right] =0 \,, \quad\ \
\left[ B, w_m \right] =  -4\,v_m-
4\,u^{IA}_m\left( \pi_I\frac{\partial}{\partial\rho^A} \right) , \quad\ \
\left[ B, u^{IA}_m \right] = -\,v_m\,
\epsilon^{AB}\left( \pi^I\frac{\partial}{\partial\rho^B} \right),
\end{equation}
\begin{equation}\label{C-com}
\begin{array}{rcl}
\left[ C_{IA}, v_m \right] & = &   \displaystyle -2\,v_m
\left( \pi_I\frac{\partial}{\partial\rho^A} \right),\\ [9pt]
\left[ C_{IA}, w_m \right] & = &  \displaystyle 4\, u_m{}_{IA} \left( \rho^B\frac{\partial}{\partial\rho^B} +4\right)
+\, 4\,u_m{}^J_{A} \left( \pi_J\frac{\partial}{\partial\pi^I}\right)
+4\,u_m{}_I^{B} \left( \rho_B\frac{\partial}{\partial\rho^A}\right) \\ [9pt]
&&  \displaystyle +\,2\,w_m \left( \pi_I\frac{\partial}{\partial\rho^A} \right) ,\\ [9pt]
\left[ C_{IA}, u^{KB}_m \right] & = &  \displaystyle
2\,u^{JC}_m
\left[ \delta_J^K \delta_C^B\left(\! \pi_I\frac{\partial}{\partial\rho^A} \!\right)
-\delta_I^K \delta_C^B\left(\! \pi_J\frac{\partial}{\partial\rho^A} \!\right)
- \delta_J^K \delta_A^B\left(\! \pi_I\frac{\partial}{\partial\rho^C} \!\right) \right] \\  [9pt]
&&  \displaystyle
+\,v_m \left[ \delta_I^K \delta_A^B\left( \!\rho^C\frac{\partial}{\partial\rho^C} \!\right)
-\delta_I^K \left(\! \rho^B\frac{\partial}{\partial\rho^A} \!\right)
- \delta_A^B\left(\! \pi^K\frac{\partial}{\partial\pi^I} \!\right) \right].
\end{array}
\end{equation}
Taking into account the relations \p{uvw-2}, \p{uvw-0} and these commutators \p{A-com}, \p{B-com}, \p{C-com}, we obtain
\begin{equation}\label{PM-2a}
\begin{array}{rcl}
\Pi^k M_{km}\Pi_l M^{lm} & = & \displaystyle r\left\{AB+BA
+\left[3\left(\! \pi^I\frac{\partial}{\partial\pi^I}\!\right)-3\left(\! \rho^A\frac{\partial}{\partial\rho^A}\!\right) -25\right]B \right.
 \\ [9pt]
& & \displaystyle  \left.\qquad\quad +\,3\epsilon^{AB}\left(\! \pi^I\frac{\partial}{\partial\rho^A} \!\right)
C_{IB} -\frac14\,\epsilon^{IJ}\epsilon^{AB}C_{IA}C_{JB}\right\}.
\end{array}
\end{equation}
In view of the equalities
\begin{equation}
\label{AB-com}
\left[A,B \right] = -\left[\left(\! \pi^I\frac{\partial}{\partial\pi^I}\!\right)-\left(\! \rho^A\frac{\partial}{\partial\rho^A}\!\right) -6\right]B \,,
\end{equation}
\begin{equation}
\label{3C}
3\epsilon^{AB}\left(\! \pi^I\frac{\partial}{\partial\rho^A} \!\right)C_{IB} =
-3\left[\left(\! \pi^I\frac{\partial}{\partial\pi^I}\!\right)-\left(\! \rho^A\frac{\partial}{\partial\rho^A}\!\right) -10\right]B \; ,
\end{equation}
expression \p{PM-2a} takes the form
\begin{equation}\label{PM-2b}
\Pi^k M_{km}\Pi_l M^{lm}  =  r\left\{
\left[2A+\left(\! \pi^I\frac{\partial}{\partial\pi^I}\!\right)-\left(\! \rho^A\frac{\partial}{\partial\rho^A}\!\right) -1\right]B
-\frac14\,\epsilon^{IJ}\epsilon^{AB}C_{IA}C_{JB}\right\},
\end{equation}
where the quantity $r$ is defined in \p{r}.
The last term in \p{PM-2b} equals
\begin{equation}\label{PM-CC}
\begin{array}{rcl}
\displaystyle -\frac14\,\epsilon^{IJ}\epsilon^{AB}C_{IA}C_{JB} & = & \displaystyle
\left[-3\left(\! \pi^I\frac{\partial}{\partial\pi^I}\!\right)+3\left(\! \rho^A\frac{\partial}{\partial\rho^A}\!\right) +5
-\left(\! \pi^I\frac{\partial}{\partial\pi^I}\!\right)\left(\! \rho^A\frac{\partial}{\partial\rho^A}\!\right)\right.
 \\ [9pt]
& & \displaystyle \left. -\,\frac12\,\epsilon^{IJ}\left(\! \pi^K\frac{\partial}{\partial\pi^I}\!\right)
\left(\! \pi_K\frac{\partial}{\partial\pi^J}\!\right)
-\,\frac12\,\epsilon^{AB}\left(\! \rho^C\frac{\partial}{\partial\rho^A}\!\right)
\left(\! \rho_C\frac{\partial}{\partial\rho^B}\!\right)\right]B \\ [9pt]
& & \displaystyle  +\,2\,\epsilon^{AB}\left(\! \rho^C\frac{\partial}{\partial\rho^A} \!\right)
\left(\! \pi^K\frac{\partial}{\partial\pi^I} \!\right)\left(\! \pi^I\frac{\partial}{\partial\rho^C} \!\right)
\left(\! \pi_K\frac{\partial}{\partial\rho^B} \!\right).
\end{array}
\end{equation}
Therefore, \p{PM-2b} is written as
\begin{equation}\label{PM-2c}
\begin{array}{rcl}
\Pi^k M_{km}\Pi_l M^{lm} & = & \displaystyle
r\left\{\left[2A-2\left(\! \pi^I\frac{\partial}{\partial\pi^I}\!\right)+2\left(\! \rho^A\frac{\partial}{\partial\rho^A}\!\right) +4
-\left(\! \pi^I\frac{\partial}{\partial\pi^I}\!\right)\left(\! \rho^A\frac{\partial}{\partial\rho^A}\!\right)\right.\right.
 \\ [9pt]
& & \displaystyle \qquad\ \ \left. -\,\frac12\,\epsilon^{IJ}\left(\! \pi^K\frac{\partial}{\partial\pi^I}\!\right)
\left(\! \pi_K\frac{\partial}{\partial\pi^J}\!\right)
-\,\frac12\,\epsilon^{AB}\left(\! \rho^C\frac{\partial}{\partial\rho^A}\!\right)
\left(\! \rho_C\frac{\partial}{\partial\rho^B}\!\right)\right]B \\ [9pt]
& & \displaystyle \qquad\ \left. +\,2\,\epsilon^{AB}\left(\! \rho^C\frac{\partial}{\partial\rho^A} \!\right)
\left(\! \pi^K\frac{\partial}{\partial\pi^I} \!\right)\left(\! \pi^I\frac{\partial}{\partial\rho^C} \!\right)
\left(\! \pi_K\frac{\partial}{\partial\rho^B} \!\right)\right\}.
\end{array}
\end{equation}

The quadratic Casimir operator of the Lorentz algebra
$\mathfrak{so}(1,5)$ with the generators $M_{mn}$
in the representation \p{M-2} has the form
\begin{equation}\label{Lor-2}
\begin{array}{rcl}
M^{mn}M_{mn} & = & \displaystyle
-\frac12 \left[\left(\! \pi^I\frac{\partial}{\partial\pi^I}\!\right) + \left(\! \rho^A\frac{\partial}{\partial\rho^A}\!\right)\right]^2
+4\left(\! \pi^I\frac{\partial}{\partial\pi^I}\!\right)-4\left(\! \rho^A\frac{\partial}{\partial\rho^A}\!\right)
\\ [9pt]
&& \displaystyle +\,4\left(\! \rho^A\frac{\partial}{\partial\pi^I}\!\right)\left(\! \pi^I\frac{\partial}{\partial\rho^A}\!\right)
 \\ [9pt]
& & \displaystyle  +\,2\left(\! \pi^K\frac{\partial}{\partial\pi^I}\!\right)
\left(\! \pi^I\frac{\partial}{\partial\pi^K}\!\right)
+2\left(\! \rho^B\frac{\partial}{\partial\rho^A}\!\right)
\left(\! \rho^A\frac{\partial}{\partial\rho^B}\!\right) .
\end{array}
\end{equation}
With the use of the relation
\begin{equation}
\label{rel-1}
\epsilon^{IJ}\left(\! \pi^K\frac{\partial}{\partial\pi^I}\!\right)\left(\! \pi_K\frac{\partial}{\partial\pi^J}\!\right)=
\left(\! \pi^J\frac{\partial}{\partial\pi^I}\!\right)\left(\! \pi^I\frac{\partial}{\partial\pi^J}\!\right)-
\left(\! \pi^I\frac{\partial}{\partial\pi^I}\!\right)^2
\end{equation}
and a similar relation for the spinor $\rho^A_\alpha$, the last two terms in \p{rel-1}
take the same form as some terms in \p{PM-2c}.

Thus, from \p{PM-2c}, \p{Lor-2}
we obtain that the six-order Casimir operator \p{C6-Cas}
(after its action
on the states $\Psi$ subject conditions
\p{eq-2} and \p{eq-1}) takes the form
\begin{equation}\label{C6-Cas-ap-fin0}
\begin{array}{rcl}
C_6 & = & \displaystyle -\, \mu^2 \left\{
-\frac14 \left(\! \pi^I\frac{\partial}{\partial\pi^I}\!\right)^2
\ +\ \frac12\left(\! \pi^K\frac{\partial}{\partial\pi^I}\!\right)\left(\! \pi^I\frac{\partial}{\partial\pi^K}\!\right)
\right.\\ [9pt]
&& \displaystyle \qquad\quad -\frac14 \left(\! \rho^A\frac{\partial}{\partial\rho^A}\!\right)^2
\ + \ \frac12\left(\! \rho^B\frac{\partial}{\partial\rho^A}\!\right)\left(\! \rho^A\frac{\partial}{\partial\rho^B}\!\right) \\ [9pt]
& & \displaystyle  \left. \qquad\quad
-\frac12 \left(\! \pi^I\frac{\partial}{\partial\pi^I}\!\right)\left(\! \rho^A\frac{\partial}{\partial\rho^A}\!\right)^{\phantom{2}} + \
\left(\! \rho^I\frac{\partial}{\partial\rho^A}\!\right)\left(\! \pi^A\frac{\partial}{\partial\pi^I}\!\right) \right\} \; .
\end{array}
\end{equation}
Here we use the fact that
the states $\Psi$ satisfy conditions
\p{eq-2} and \p{eq-1}, which respectively implies
the relations $r \Psi=2\mu^2\Psi$ and $B \Psi=\Psi/2$.
The last expression for $C_6$
given in \p{C6-Cas-ap-fin0} is represented in the
form
\begin{equation}
\label{C6-ap}
C_6 \ = \, -\, \mu^2 J_aJ_a \; ,
\end{equation}
where we introduce the $\mathrm{SU}(2)$ generators
$J_a$ defined in \p{Ta-def}.

\end{document}